\documentclass[aps,prx,twocolumn,nofootinbib,superscriptaddress]{revtex4-1}

\usepackage{graphicx}
\usepackage{amssymb}
\usepackage{amsmath}
\usepackage{color}
\usepackage{physics}
\usepackage{bbm}
\usepackage{hyperref}
\usepackage{epstopdf}
\usepackage{dsfont}

\usepackage{enumerate}
\usepackage{kantlipsum}
\usepackage{refcount}

\newcommand{\be}{\begin{equation}}
\newcommand{\ee}{\end{equation}}
\newcommand{\bs}{\begin{split}}
\newcommand{\es}{\end{split}}

\newcommand{\OTOC}{\text{OTOC}}
\newcommand{\TOC}{\text{TOC}}

\newcommand{\err}{\varepsilon}

\newcommand{\Jl}{J_{\ell}}
\newcommand{\J}{J}

  \renewcommand\[{\ensuremath \left[}
  
\def\:={\,\raisebox{0.85pt}{.}\hspace{-2.78pt}\raisebox{2.85pt}{.}\!\!=\,}
\def\=:{\,=\!\!\raisebox{0.85pt}{.}\hspace{-2.78pt}\raisebox{2.85pt}{.}\,}

\newcommand{\FigOne}{
\begin{figure}
\centering
\includegraphics[width=\columnwidth]{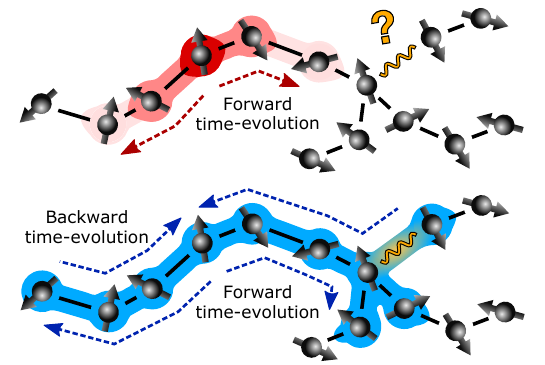}
\caption{
Schematic of time-ordered correlators (TOCs) and out-of-time-order correlators (OTOCs) in strongly-interacting systems.
TOCs typically decay in $O(1)$ times and distances (top, red), making it hard to learn features (yellow bond) that manifest only at late times.
OTOCs utilize backwards time-evolution to ``refocus'' many-body correlations (bottom, blue), enabling learning of such features.
} 
\label{fig: 1}
\end{figure}
}

\newcommand{\FigProbe}{
\begin{figure}
\centering
\includegraphics[width=0.75\columnwidth]{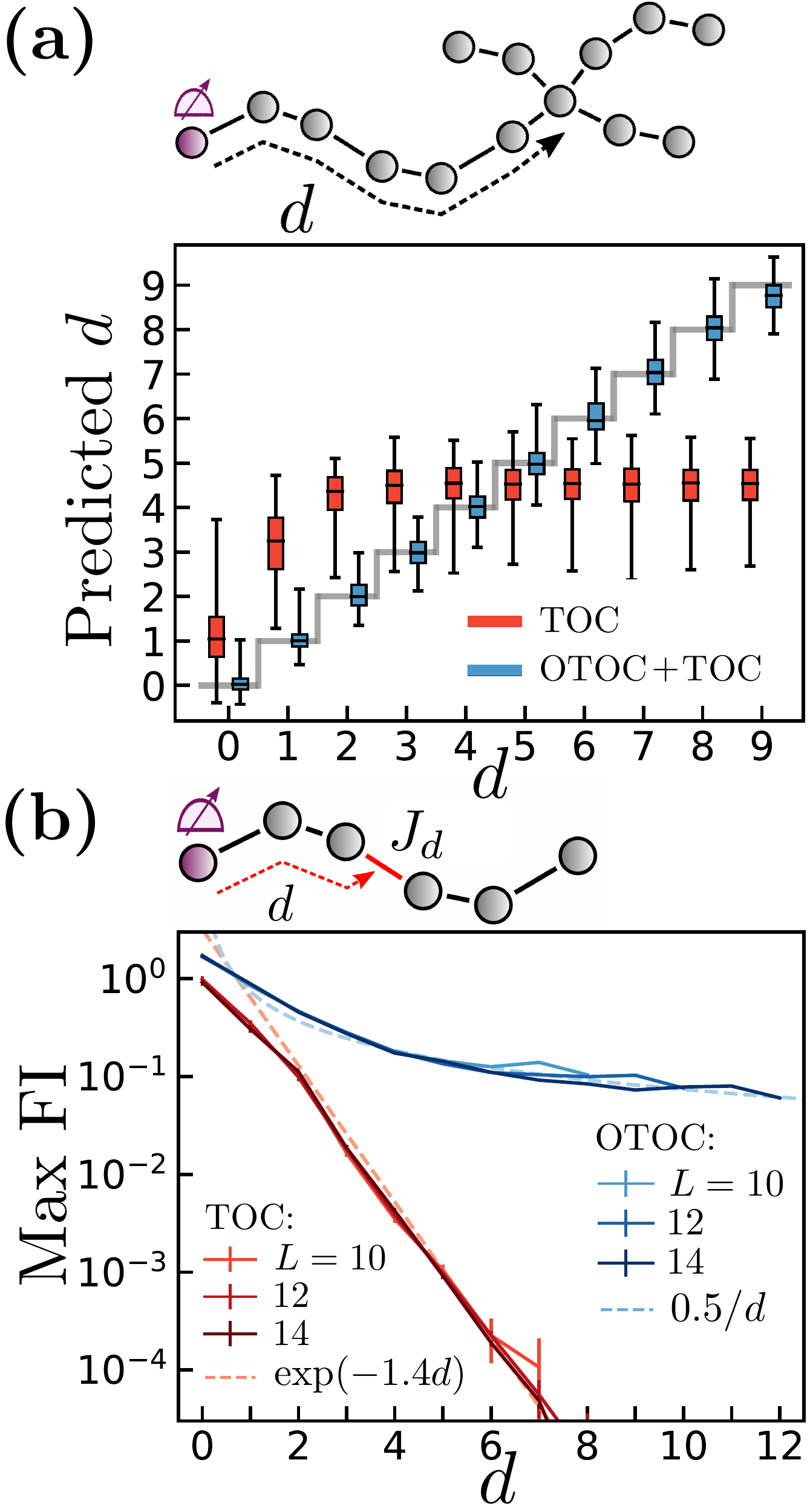}
\caption{
Learning with state preparation and read-out restricted to a probe qubit, and local unitary control over the remaining system.
\textbf{(a)} Results from SVM regression for learning the distance, $d$, in the spin geometry shown, with access to TOCs (red) or both TOCs and OTOCs (blue). Color bars (black ticks) denote $75\%$ ($100\%$) percentiles of predictions on $200$ disorder realizations, and grey step function represents the actual $d$.   
\textbf{(b)} Fisher information, $\text{FI}(J_d | C)$, of an interaction, $J_d$ (top; red line), a distance $d$ away from the probe (top; purple circle), maximized over all correlators, $C$, in an $L$-qubit 1D chain. 
The FI decays exponentially in $d$ when $C$ is time-ordered (red), and algebraically, $\sim 1/d$, when $C$ is out-of-time-order (blue).
} 
\label{fig: probe}
\end{figure}
}

\newcommand{\FigWeak}{
\begin{figure}
\centering
\includegraphics[width=0.75\columnwidth]{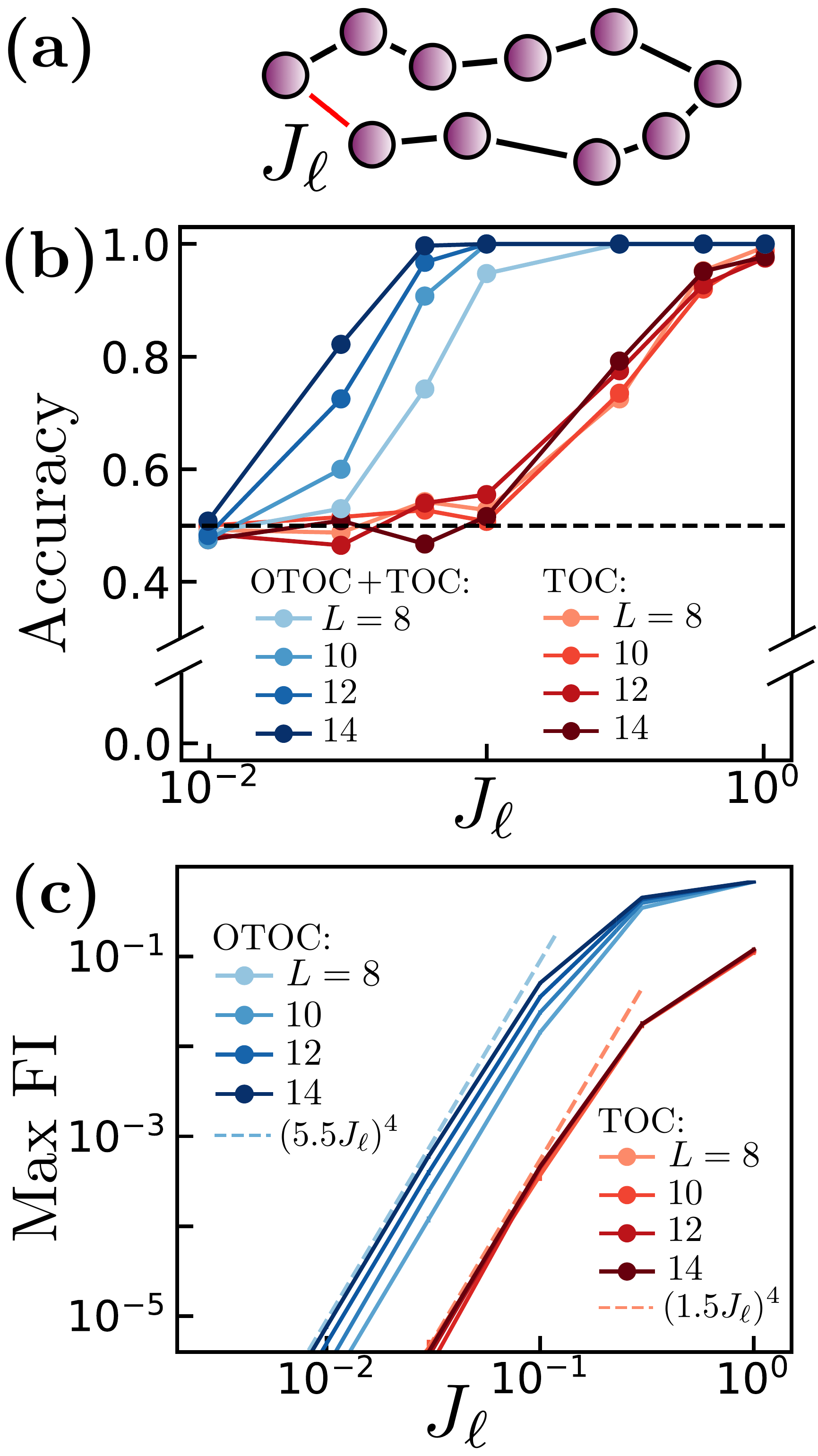}
\caption{
\textbf{(a)} Learning a weak ``link'' interaction (red line) in a 1D spin chain with otherwise strong interactions (black lines).
\textbf{(b)} Accuracy of binary SVM classification of whether the link is present or absent, as a function of the link strength $\Jl$ and at fixed read-out error $\delta = 3\%$.
Learning via OTOCs can detect smaller $\Jl$ as $L$ increases, while the TOC can only detect relatively large $\Jl$, independent of $L$.
\textbf{(c)} The maximum Fisher information $\text{FI}(\log(J_\ell) | C)$ of $J_\ell$ decays $\sim \! J_\ell^4$ for small $J_\ell$, and is enhanced in OTOCs (blue) compared to TOCs (red) by a factor that increases with $L$.
} 
\label{fig: weak}
\end{figure}
}

\newcommand{\FigError}{
\begin{figure}
\centering
\includegraphics[width=0.75\columnwidth]{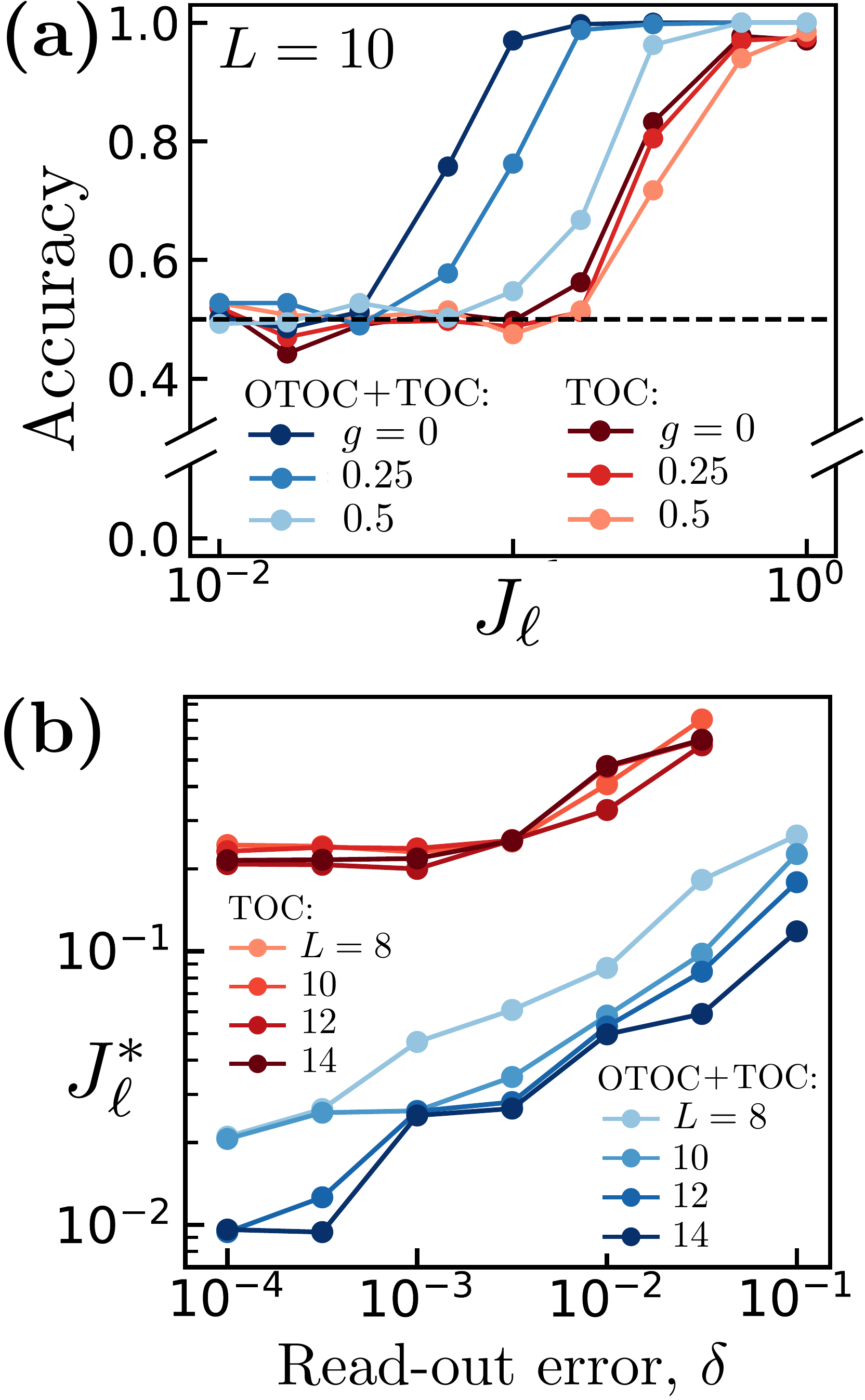}
\caption{
Learning as a function of experimental error, in the ``weak interaction'' learning task of Fig.~\ref{fig: weak}(a).
\textbf{(a)} Accuracy of binary SVM classification as in Fig.~\ref{fig: weak}(a), now with a coupling $g$, to an extrinsic cavity mode that is not time-reversed (cavity frequency $\omega = 1.7$). Despite imperfect time-reversal, learning via OTOCs continues to provide an advantage up to large spin-cavity couplings $g \sim 0.5$.
\textbf{(b)} The minimum link strength $J_\ell^*$ classifiable with $>90\%$ accuracy as a function of read-out error $\delta$, obtained by repeating Fig.~\ref{fig: weak}(a) for each $\delta$. The minimum link strength in general decreases with decreasing $\delta$; for learning via TOCs, this decrease plateaus for $\delta \lesssim 0.1\%$, indicating that learning below this value is not limited by read-out error.
} 
\label{fig: error}
\end{figure}
}

\newcommand{\FigToy}{
\begin{figure}
\centering
\includegraphics[width=0.85\columnwidth]{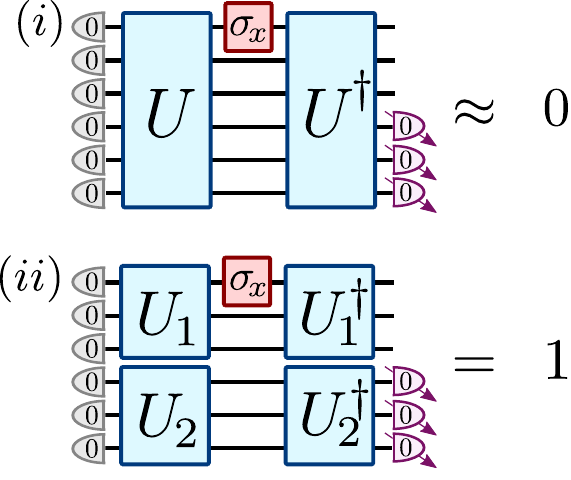}
\caption{
Solution to the disjoint unitary problem with out-of-time-order measurements. 
The state $\ket{0}^{\otimes n}$ is prepared and the unknown unitary (either $U$ or $U_1 \otimes U_2$) is applied.  Next $\sigma_x$ is applied to the first qubit, followed by the inverse of the unknown unitary.  Finally, it is checked if the second block of $n/2$ qubits ends up in the all zero state.  If so, then the hidden unknown unitary is $U_1 \otimes U_2$ as per case (\textit{ii}); if not, then the unknown unitary is $U$ as per case (\textit{i}).}
\label{fig:toy}
\end{figure}
}

\newcommand{\FigExtrapolate}{
\begin{figure}
\centering
\includegraphics[width=0.7\textwidth]{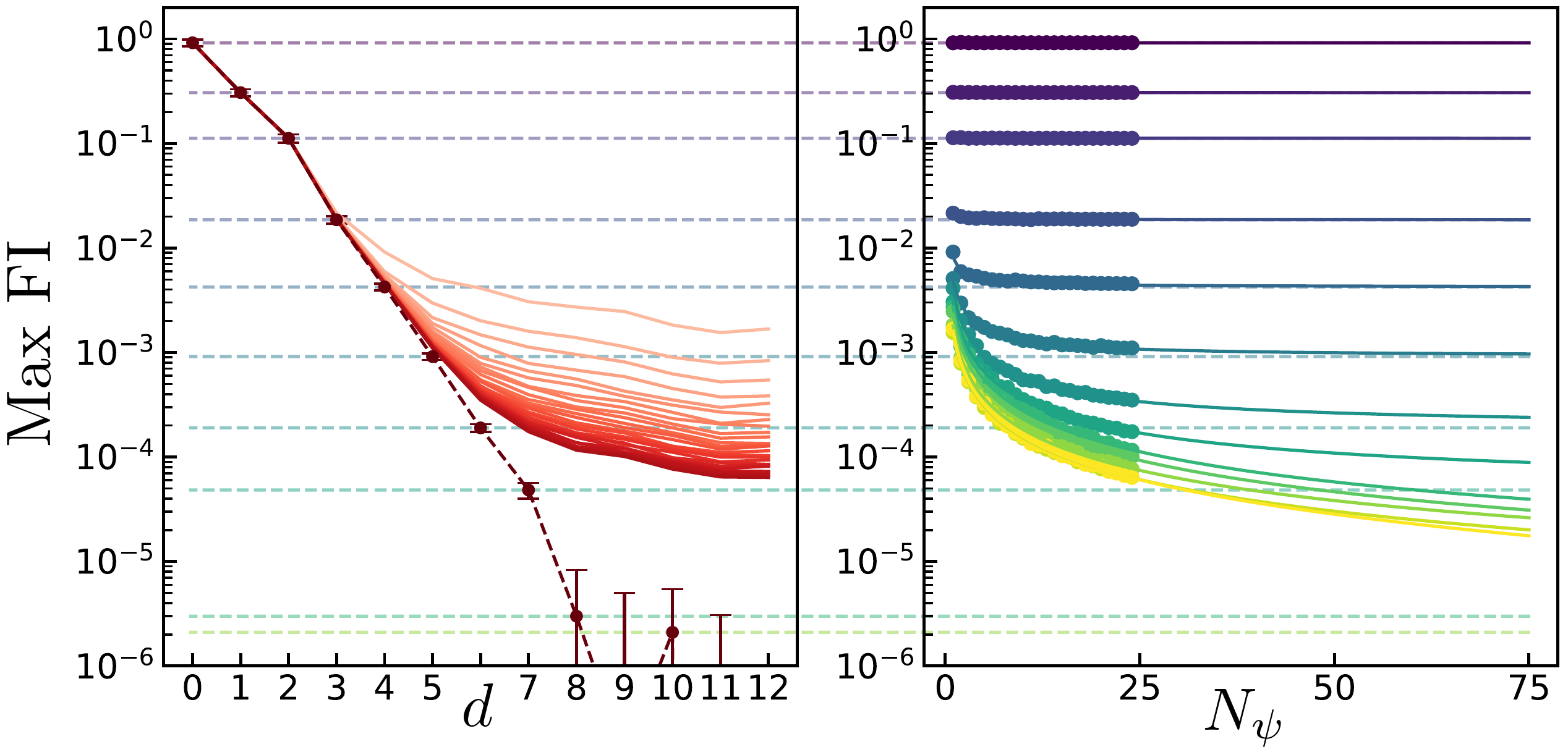}
\caption{
Depiction of the extrapolation method used to calculate the maximum Fisher information over time-ordered correlators [Fig.~\ref{fig: probe}(b)].
Each correlation function is computed for 25 Haar-random values of the state $\ket{\psi}$ [Eq.~(\ref{eq: corr fns Haar random})].
For each value of $N_\psi$ between 1 and 25, we choose a random subset of $N_\psi$ values of $\ket{\psi}$ and compute the average correlation function over the subset.
(Left) For each value of $N_\psi$, we then compute the maximum Fisher information over all correlation functions, $\text{max FI}(N_\psi)$ (solid red lines, darker lines corresponds to higher $N_\psi$).
(Right) Our estimate of the maximum Fisher information at infinite temperature (dotted lines, both plots) is obtained by fitting $\text{max FI}(N_\psi) = \text{max FI}(\infty) + A/N_\psi$ and taking $N_\psi \rightarrow \infty$ (points denote data, solid lines denote $1/N_\psi$-fit).
} 
\label{fig: extrapolate}
\end{figure}
}

\newcommand{\FigHamProbe}{
\begin{figure}
\centering
\includegraphics[width=0.7\textwidth]{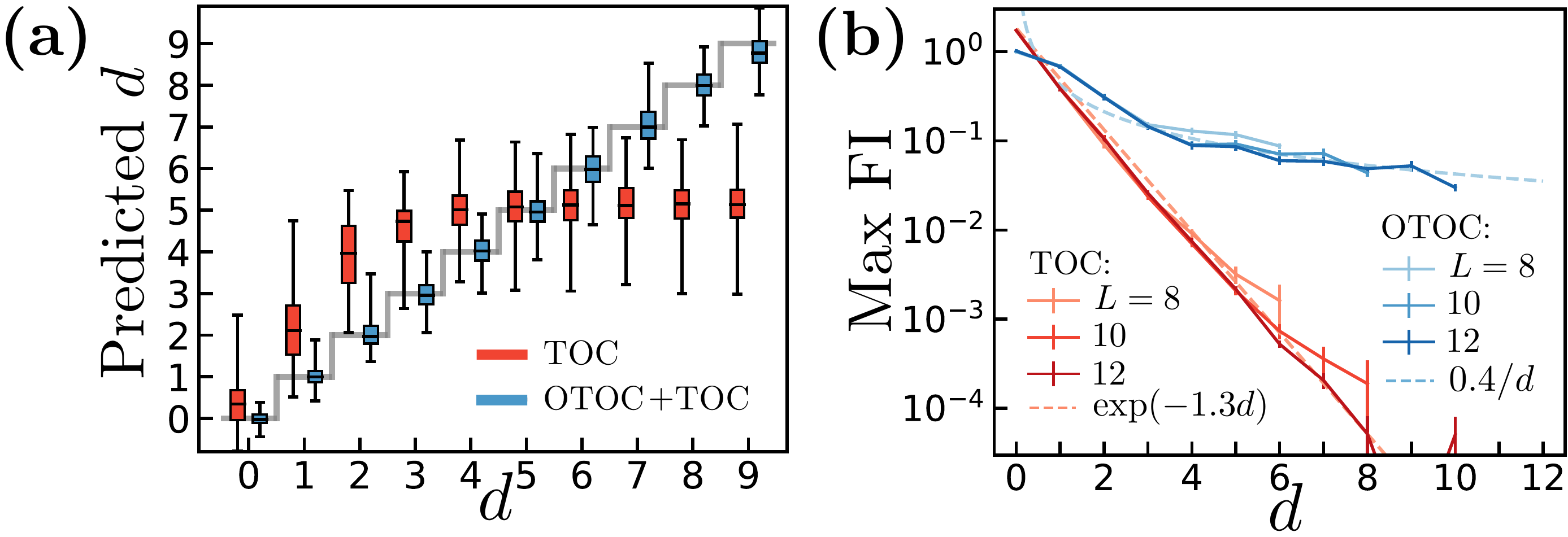}
\caption{
Learning in the restricted access scenario under Hamiltonian evolution. Numerical simulations are performed identically to Fig.~\ref{fig: probe} but now with Hamiltonian evolution under $(H_c + H_f)/2$ instead of Floquet evolution. 
In both $\textbf{(a)}$ the learning task and $\textbf{(b)}$ the Fisher information, the results for learning Hamiltonian dynamics are qualitatively similar to the results for learning Floquet dynamics [Fig.~\ref{fig: probe}].
In $\textbf{(b)}$, the maximum Fisher information is averaged over 100 disorder realizations for both TOCs and OTOCs.
At large $d$, we expect the Fisher information for Hamiltonian evolution to approach a power law decay $\sim \! 1/d^4$ (see Appendix~\ref{sec: Fisher restricted}), but this cannot be observed in our finite-size numerics.
} 
\label{fig: probe ham}
\end{figure}
}

\begin{document}

\author{Thomas Schuster}
\email{tsschuster@google.com}
\affiliation{Google Quantum AI, Venice, CA 90291, USA}
\affiliation{Department of Physics, University of California, Berkeley, CA 94720 USA}
\author{Murphy Niu}
\affiliation{Google Quantum AI, Venice, CA 90291, USA}
\author{Jordan Cotler}
\affiliation{Society of Fellows, Harvard University, Cambridge, MA, USA}
\affiliation{Black Hole Initiative, Harvard University, Cambridge, MA, USA}
\author{Thomas O'Brien}
\affiliation{Google Quantum AI, Venice, CA 90291, USA}
\author{Jarrod R. McClean}
\affiliation{Google Quantum AI, Venice, CA 90291, USA}
\author{Masoud Mohseni}
\email{mohseni@google.com}
\affiliation{Google Quantum AI, Venice, CA 90291, USA}

\date{\today}

\begin{abstract}
Learning the properties of dynamical quantum systems underlies applications ranging from nuclear magnetic resonance spectroscopy to quantum device characterization.
A central challenge in this pursuit is the learning of strongly-interacting systems, where conventional observables decay quickly in time and space, limiting the information that can be learned from their measurement.
In this work, we introduce a new class of observables into the context of quantum learning---the out-of-time-order correlator---which we show can substantially improve the learnability of strongly-interacting systems by virtue of displaying informative physics at large times and distances.
We identify two general scenarios in which out-of-time-order correlators provide a significant advantage for learning tasks in locally-interacting systems: ($i$) when experimental access to the system is spatially-restricted, for example via a single ``probe'' degree of freedom, and ($ii$) when one desires to characterize weak interactions whose strength is much less than the typical interaction strength.
We numerically characterize these advantages across a variety of learning problems, and find that they are robust to both read-out error and decoherence. 
Finally, we introduce a binary classification task that can be accomplished in constant time with out-of-time-order measurements.   In a companion paper~\cite{cotler2022information}, we prove that this task is exponentially hard with any adaptive learning protocol that only involves time-ordered operations.
\end{abstract}

\title{Learning quantum systems via out-of-time-order correlators}
\date{\today}
\maketitle


\section{Introduction}

Learning properties of quantum systems can pose challenges not present in their classical counterparts~\cite{mohseni2008quantum,huang2022}.
These differences often stem fundamentally from the existence of entanglement---measurements of a quantum system that is highly entangled with another system or the environment reveal little information from which to learn.
In practical settings, these difficulties are most commonly encountered in \emph{strongly-interacting} quantum systems.
Strong interactions can introduce non-local entanglement throughout the system at short time scales, and are found to thereby inhibit the learning of system properties (e.g.~the Hamiltonian) from physical observables~\cite{wiebe2014quantum,aharonov2021quantum,huang2021quantum,zhou2020quantum,o2021quantum}.

The ubiquity of strong interactions in experimental applications of quantum learning has spurred a variety of solutions to this problem.
For instance, in nuclear magnetic resonance (NMR) spectroscopy, a suite of technologies have been developed to controllably dampen undesired strong interactions between solid-state nuclear spins, which has enabled the identification of hitherto inaccessible molecular structures~\cite{laws2002solid}.
In a similar spirit, in quantum device characterization~\cite{wang2015hamiltonian} and quantum sensing~\cite{zhou2020quantum}, dynamical decoupling control sequences~\cite{ezzell2022dynamical} can effectively eliminate unwanted interactions and improve learning of the residual interactions.
Other approaches include learning by transducing quantum data from the system onto a quantum simulator~\cite{wiebe2014quantum,aharonov2021quantum,huang2021quantum,wang2017experimental,huang2022}, or learning from high-precision local measurements at early times, before entanglement has formed~\cite{bairey2019learning,haah2021optimal}.
Nonetheless, owing to incomplete control or limited experimental precision, many physical systems remain unlearnable with existing approaches.

\FigOne


In this paper, we introduce a different paradigm for learning in strongly-interacting quantum systems---learning via out-of-time-order correlators (OTOCs).
First studied in early works on semi-classical methods~\cite{larkin1969quasiclassical} and NMR~\cite{baum1985multiple}, the OTOC has more recently initiated a renaissance of work at the intersection of quantum information theory, many-body dynamics, and quantum gravity (e.g.~\cite{shenker2014black, stanford2016many, maldacena2016bound, maldacena2016remarks, hosur2016chaos, cotler2017chaos, cotler2018out, nahum2018operator, von2018operator}).
Physically, OTOCs quantify the spread of local quantum information into highly non-local, entangled correlations~\cite{xu2022scrambling}.
Experimental measurements of the OTOC typically employ \emph{reversed} time-evolution to refocus these correlations, and have been performed on dozens of qubits in superconducting quantum processors and trapped ion quantum simulators, and hundreds of nuclear spins in NMR spectroscopy~\cite{garttner2017measuring,mi2021information,sanchez2021emergent,dominguez2021decoherence}.

In this work, we utilize the OTOC as a \emph{tool} for learning properties of strongly-interacting quantum systems.
Our application is motivated by a simple intuition: while time-ordered observables decay quickly as a system becomes entangled, out-of-time-order observables continue to fluctuate up to long times (Fig.~\ref{fig: 1}).
Guided by this intuition, we demonstrate the power of learning via OTOCs across a range of physical systems, supported by numerical studies, phenomenological estimates, and rigorous information-theoretic proofs.
We begin in locally-interacting systems, where we identify two general scenarios in which OTOCs provide a strong learning advantage: ($i$) when experimental access to the system is spatially-restricted, for example via a single ``probe'' qubit~\cite{burgarth2009coupling,lovchinsky2016nuclear,gentile2021learning}, and ($ii$) for detecting weak interactions in an otherwise strongly-interacting system~\cite{wang2015hamiltonian,zhou2020quantum}.
We characterize these advantages using both information-theoretic measures (the Fisher information) and performance metrics for concrete learning tasks.
Moreover, we find that the advantages are robust to experimental read-out error and time-reversal imperfections arising from strong coupling with an environment or decoherence.
Finally, motivated by recent advances in \emph{provable} learning advantages~\cite{aharonov2021quantum, chen2021exponential, huang2021quantum}, we introduce a learning task involving distinguishing two classes of unitary operations if given oracle access.
In a companion work~\cite{cotler2022information}, we prove that OTOCs provide an exponential advantage in performing this task over any time-ordered learning protocol.


\section{Behavior of time-ordered vs. out-of-time-order correlators}

\FigProbe

We begin by reviewing the phenomenology of time-ordered and out-of-time-order correlators in ergodic locally-interacting systems (Fig.~\ref{fig: 1}).
A time-ordered correlator (TOC) is defined as any correlation function that takes the following general form:
\begin{equation} \label{eq: TOC generic def}
C_{\TOC} = \tr( A_k(t_k) \ldots A_1(t_1) \, \rho \, B_1(t'_1) \ldots B_\ell(t'_\ell) ).
\end{equation}
where the operators $A, B$ increase in time away from the initial density matrix $\rho$, i.e. $t_k > \cdots > t_1$ and $t'_\ell > \cdots > t'_1$.
Time-ordered correlators can be measured by evolving the state $\rho$ forward in time (e.g.~via Hamiltonian evolution $O(t) = e^{iHt} O e^{-iHt}$) while applying intermediary quantum operations at each time $t_i, t'_j$~\cite{fn4}.
Any correlation function that does not obey this form is called an \emph{out-of-time-order} correlator.

A common example of a time-ordered correlator is the two-point function,
\begin{equation} \label{eq: TOC def}
C_{\TOC} = \expval{ V_{x}(t) W_{x'}(0) },
\end{equation}
where $\langle \cdot \rangle \equiv \tr(\cdot)/2^L$ denotes the infinite temperature trace for $L$ qubits, and $V_x, W_{x'}$ are local operators at sites $x, x'$.
Such correlators measure the spread of local quantities in space and time; for instance, how much spin prepared at site $x'$ at time zero has transferred to site $x$ at time $t$.
A wide range of literature on thermalization in strongly-interacting systems has found that local TOCs typically decay quickly, i.e.~in $O(1)$ times, to their thermal values~\cite{d2016quantum}.
This quick decay can inhibit learning tasks, since no additional information can be acquired from the TOC at times after the decay has occurred~\cite{o2021quantum}.

Meanwhile, the prototypical out-of-time-order correlator is the four-point function~\cite{xu2022scrambling},
\begin{equation} \label{eq: OTOC def}
C_{\OTOC} = \expval{ V_{x}(t) W_{x'}(0) V^\dagger_{x}(t) W_{x'}(0)},
\end{equation}
with local operators $V_x, W_{x'}$.
Unlike time-ordered measurements, OTOCs typically require both forwards \emph{and} backwards time-evolution to measure~\cite{xu2022scrambling,cotler2022information}.
(Importantly for our application, nearly all experimental techniques for time-reversal rely only on the type of interaction being reversed and require no knowledge of the specific Hamiltonian, which one might wish to learn. For example, the same pulse sequence reverses an arbitrary dipole-dipole coupling Hamiltonian in an NMR experiment~\cite{sanchez2021emergent}.)
Physically, the OTOC probes whether information encoded at site $x'$ at time zero is contained in correlations involving site $x$ at time $t$. This is quantified by the squared commutator of a time-evolved operator at $x$ with a local operator at $x'$, $\expval{\left| \left[  V_{x}(t), W_{x'}(0) \right]\right|^2} = 1 - C_{\OTOC}$.
In local strongly-interacting systems, operators are expected to spread ballistically according to the connectivity of the system~\cite{lieb1972finite,aleiner2016microscopic,nahum2018operator}.
Crucially, this spread continues for a duration  proportional to the system's spatial extent $\sim L$ by which time the information has been delocalized across the entire system.

This phenomenology leads to two central intuitions for learning from OTOCs.
First, the dynamics of the OTOC contain information primarily about the connectivity of the system under study.
Second, the OTOC continues to reveal such information up to $O(L)$ times, long after TOCs have decayed. Notice that this timescale increases as the system size increases. 
In what follows, we apply these intuitions to identify two broad regimes where access to OTOCs provides a significant learning advantage.


\section{Learning with restricted access}

The first regime we consider is learning in systems with \emph{restricted access}.
Specifically, motivated by recent advances in solid-state defects~\cite{boss2016one,lovchinsky2016nuclear,gentile2021learning} and NMR~\cite{theis2016direct,blanchard2021lower,burgarth2017evolution}, we focus on the scenario where an experimenter has state preparation and read-out capabilities over only a single ``probe'' qubit interacting with a larger system that one wishes to learn.
We note that high-fidelity OTOC measurements have already been achieved in similar setups by using rapid global pulse sequences to reverse time-evolution~\cite{wei2018exploring,sanchez2019emergent,sanchez2021emergent}.
Previous theoretical approaches to learning in this scenario have been limited to non-interacting dynamics~\cite{burgarth2009coupling,burgarth2009indirect,di2009hamiltonian,zhang2014quantum,burgarth2017evolution,sone2017hamiltonian}.
Meanwhile, experiments have found that it is in general difficult to learn features of a system that are distant from the probe qubit~\cite{lovchinsky2016nuclear,gentile2021learning}.
In strongly-interacting systems, this difficulty can be understood from the quick decay of correlation functions in space and time.
Here, we provide evidence via phenomenological estimates (Appendix~\ref{sec: Fisher restricted}) and numerical simulations (Fig.~\ref{fig: probe}) that access to OTOCs can exponentially improve the learnability of distant features.

To be concrete, we will assume for now that the experimenter has local unitary control over the qubits of the larger system~\cite{fn1}.
We will also assume that the larger system begins in an infinite temperature (i.e.~maximally mixed) state, which is the natural scenario in NMR and solid-state defect setups~\cite{boss2016one,theis2016direct}.
Within these assumptions, a simple class of measurement protocols proceeds as follows:
\begin{enumerate}
\item Prepare the probe qubit $p$ in an eigenstate of an operator $V_p$, such that the density matrix of the entire system is $\rho = \frac{1}{2}(\mathbbm{1}_p+V_p) \otimes \frac{1}{2^{L-1}}\,\mathbbm{1}_{\text{sys}}$.
\item Time-evolve by time $\tau$.
\item Perturb the system by a unitary operation $W_x$ on a qubit $x$.
\item Time-evolve by a time $\tau'$.
\item Read out the expectation value of $V_p$ on the probe qubit.
\end{enumerate}
Taking $\tau, \tau'$ to be positive (e.g. $\tau = \tau' = t/2$), this allows measurement of time-ordered correlation functions of the form $\langle V_p(t) \, W_x(t/2) \, V_p(0) \, W^\dagger_x(t/2) \rangle$.
With access to reversible time-evolution (e.g.~$\tau = -\tau' = t$), the above protocol also allows measurement of out-of-time-order correlation functions $\langle V_p(0) \, W_x(t) \, V_p(0) \, W^\dagger_x(t) \rangle$. 
In Appendix~\ref{sec: Fisher restricted} and~\ref{sec: additional numerics}, we discuss how learning is modified when $W$ is instead a \emph{global} spin rotation over the larger system.

\FigWeak

We begin our exploration of learning via OTOCs by introducing a concrete learning task.
We consider the following scenario: one is given access to a quantum system consisting of two spin chains intersecting at a distance $d$ from a probe qubit [Fig.~\ref{fig: probe}(b)].
The value of $d$ as well as the specific Hamiltonian parameters of the system are unknown (see below for the specific distribution that the Hamiltonian is drawn from).
The goal is to learn the value of $d$, i.e.~the geometry of the system, from measurements of the system's correlation functions.

To solve this task, we assume that the experimenter is capable of simulating quantum dynamics on either a classical or quantum computer.
Since the task involves high-dimensional input data (i.e.~the correlators for every $x,t$), we will approach it using machine learning techniques.
Specifically, we envision using the quantum simulator to compute the correlation functions of an ensemble of Hamiltonians for each value of $d$.
These ensembles can then be used to train a classical learning model to predict an unknown Hamiltonian's value of $d$ given its correlation functions.

Let us briefly summarize our numerical simulations in more detail (see Appendix~\ref{sec: numerical} for a complete description).
Throughout this work, we consider spin systems with disordered on-site fields, $H_f = \sum_{i,\alpha} h_i^\alpha \sigma^\alpha_i$ with $h_i^\alpha \in [-1,1]$ and $\alpha = x,y,z$, and dipolar interactions between neighboring spins, $H_c = \sum_{\langle i j \rangle} J_{ij} (\sigma^x_i \sigma^x_j+\sigma^y_i \sigma^y_j-2\sigma^z_i \sigma^z_j)$ with $J_{ij} \in [0.6,1.4]$.
We specify to Floquet dynamics consisting of alternating applications of $H_f$ and $H_c$ for time $T = \pi/2$, and simulate time-evolution via Krylov-subspace methods~\cite{dynamite}.
We expect that learning Floquet dynamics will be qualitatively similar to learning time-independent Hamiltonian dynamics at moderate times and distances, which we are restricted to in our numerics (see Appendix~\ref{sec: additional numerics} for numerical support of this statement).
At larger distances we expect Hamiltonian dynamics to be dominated by hydrodynamics of the conserved energy (Appendix~\ref{sec: Fisher restricted}) and the two will differ.

Returning to the learning task at hand, we train a support vector machine (SVM) on 3000 randomly drawn Hamiltonians (300 for each value of $d = 0,\ldots,9$), and test its performance on 2000 additional Hamiltonians.
To ensure that learning is not sensitive to fine-tuned features of the correlation functions, we add a Gaussian distributed ``read-out error'' to all correlation functions, with mean zero and standard deviation $\delta = 3\%$.
The model's predictions as a function of the actual value of $d$ are displayed in Fig.~\ref{fig: probe}(b), for learning either via TOCs (red) or both TOCs and OTOCs (blue).
We find that learning via OTOCs allows accurate predictions of $d$ within $\pm 1$ of its actual value for all distances probed (up to $d = 9$).
In contrast, with access to only TOCs, the model performs significantly worse for all $d$ and resorts to nearly random guessing for $d \gtrsim 3$.

To evaluate the learning advantage of OTOCs independent of a specific learning task, we turn to the Fisher information (FI).
The FI quantifies the amount of information that a random variable (e.g.~a correlation function $C$, measured within some read-out error $\delta$) carries about an unknown parameter (e.g. a coupling strength, $J$), and thereby bounds the ultimate learnability of the parameter~\cite{kay1993fundamentals}.
If one assumes that read-out errors are normally distributed, the FI is simply a squared derivative, $\text{FI}(J | C) \equiv \delta^2 \text{FI}( J | C ; \delta ) = \left| \partial C / \partial J \right|^2$, where we remove the $\delta$-dependence by introducing a factor $\delta^2$.

We numerically compute the FI in ergodic 1D spin chains, where one seeks to learn a coupling $J_d$ lying a distance $d$ away from a probe qubit [Fig.~\ref{fig: probe}(b) inset]~\cite{burgarth2009coupling,burgarth2009indirect,di2009hamiltonian,zhang2014quantum,burgarth2017evolution,sone2017hamiltonian}.
We consider the same set of correlation functions as specified for the learning task in Fig.~\ref{fig: probe}(a).
In Fig.~\ref{fig: probe}(b), we plot the maximum Fisher information $\max_C \text{FI}(J | C)$ over all correlation functions (i.e.~over all $x, t$), averaged over 200 and 1000 disorder realizations for TOCs and OTOCs respectively.
We find that the maximum FI of TOCs (red) decays exponentially in the distance $d$ from the probe qubit.
In contrast, the maximum FI of OTOCs (blue) follows a slow algebraic decay, $\sim 1/d$, thereby achieving a multiple-order-of-magnitude advantage over TOCs even at modest distances, $d \gtrsim 3$. 
This algebraic decay arises from the $\sim \! \sqrt{t}$ broadening of the OTOC wavefront in time~\cite{nahum2018operator}, see Appendix~\ref{sec: Fisher restricted} for a full phenomenological derivation.


\section{Learning weak interactions}

We now turn to our second learning scenario: characterizing weak interactions in an otherwise strongly-interacting system.
Such characterization is notoriously difficult because weak interactions take long times to manifest (of order the inverse interaction strength), at which point TOCs have decayed due to the strong interactions.
Previous approaches require either dynamical decoupling of the strong interactions~\cite{wang2015hamiltonian,zhou2020quantum} or high-precision measurements at early times~\cite{bairey2019learning,haah2021optimal}.
We will now show that access to OTOCs allows one to side-step these requirements when characterizing weak interactions that \emph{change the connectivity} of a strongly-interacting system.
%
Notably, in contrast to the previous learning scenario, this advantage holds when the experimenter is capable of measuring all local correlation functions of the system of interest.

For concreteness, we specialize to 1D spin chains with a single ``weak link'' interaction, of strength $\Jl$ much less than the typical interaction strength $J$ [see Fig.~\ref{fig: weak}(b) inset].
We consider TOCs and OTOCs of the form Eq.~\eqref{eq: TOC def} and Eq.~\eqref{eq: OTOC def}, where $x, x'$ run over all qubits in the system.
We anticipate that access to more general correlators within a given time-ordering, e.g. via shadow tomography or related techniques~\cite{aaronson2019shadow, cotler2020quantum, huang2020predicting}, will not qualitatively change the observed physics (see Appendix~\ref{sec: numerical}).

\FigError

We begin as before with a concrete learning task.
Specifically, we suppose that one is given access to a spin chain with unknown Hamiltonian parameters and either no link interaction ($\Jl \rightarrow 0$) or a fixed non-zero weak link interaction strength $\Jl$. 
For each fixed value of $\Jl$, we train a binary SVM classifier on the correlation functions [Eqs.~(\ref{eq: TOC def}),~(\ref{eq: OTOC def})] of 300 disorder samples, again including a read-out error $\delta = 3\%$ in each correlator value.
We test model performance on 200 additional samples; the resulting classification accuracies are shown in Fig.~\ref{fig: weak}(b).
We observe the following general trends: ($i$) the accuracy decreases as $\Jl$ decreases; ($ii$) learning via both OTOCs and TOCs (blue) allows detection of $\sim \! 10$ times smaller $\Jl$ than learning via only TOCs (red); and ($iii$) OTOCs allow detection of increasingly small $\Jl$ as the size $L$ of the chain increases.

To understand this behavior analytically, we first note that the optimal correlation functions for detecting the link will typically involve operators lying immediately adjacent to that link, on both of its sides.
These correlators measure either the transfer of spin polarization (for TOCs) or operator support (for OTOCs) across the link, and will be non-trivial only if the link interaction strength is nonzero.
For TOCs, one expects spin polarization to cross the link incoherently, at a rate $\sim \! \Jl^2/\J$, where $J$ is the typical strong interaction strength.
Combined with an overall exponential decay of spin in time (if the system has no conserved quantities), we expect $C_\TOC \sim (\Jl^2 / \J) \,t\, e^{-\J t}$.
For OTOCs, one expects an operator's support to cross the link at a similar rate, $1-C_\OTOC \sim (\Jl^2 / \J) t$.
Crucially however, this growth persists until much \emph{later} times, $t \sim L/\J$, at which information traveling ``around'' the chain will abruptly cause the OTOC to decay to zero.
The optimal time for detecting the link occurs when these correlators are maximized, since each is zero in the absence of the link.
The TOC is maximized at an order one time $t \sim 1/\J$, at which the correlator magnitude $C_\TOC \sim \Jl^2 / \J^2$ is suppressed by the square of the weak link interaction strength.
In contrast, the OTOC is maximized at a much later time $t \sim L/\J$, and thereby features a magnitude $1 - C_\OTOC \sim L (\Jl^2 / \J^2)$.
In both cases we see that detection of the link becomes more difficult as the link strength decreases.
Detection via the OTOC is enhanced by a factor of $L$, which captures the connectivity change associated with the link.

We confirm these estimates quantitatively by computing the Fisher information of the link interaction strength.
In Fig.~\ref{fig: weak}(b), we plot the maximum Fisher information $\max_C \text{FI}(\log(J_\ell) | C)$ over all local correlation functions, averaged over 100 disorder realizations.
Here, we consider the logarithm of the link interaction strength in order to appropriately compare the Fisher information over multiple orders of magnitude of the interaction. The Fisher information of $\log(\Jl)$ bounds the learnability of the interaction strength as a \emph{percentage} of its actual value.
Applying our phenomenological estimates, we predict that $\text{FI} \sim \Jl^4/J^4$ for TOCs, and $\text{FI} \sim L^2 \Jl^4/J^4$ for OTOCs.
Observing Fig.~\ref{fig: weak}(b), we indeed find that the FI is suppressed by $\sim \! \Jl^4$ (dashed lines) for small $\Jl$, and displays a multiplicative advantage for OTOCs (blue) compared to TOCs (red), which grows as $L$ increases.

\section{Effect of experimental errors}

Let us now address the impact of experimental errors on learning. 
We begin with errors that accumulate throughout time-evolution.
These may occur from extrinsic decoherence or imperfect time-reversal dynamics, each of which disrupt the non-local correlations probed by the OTOC~\cite{swingle2018resilience,mi2021information,sanchez2021emergent,dominguez2021decoherence,schuster2022time}.
While this disruption can be mitigated via independent error estimates~\cite{swingle2018resilience,mi2021information}, for sufficiently large errors these estimates involve measuring quantities of small magnitude (comparable to the TOC), squandering the OTOC's learning advantage.
In Appendix~\ref{sec: Fisher restricted} we estimate that our previous results are modified in the presence of a small local error rate $\err \ll J$ as follows: in the first learning regime, the OTOC maintains its advantage up to distances $d \lesssim J/\err$; in the second regime, the $L$-fold advantage is replaced by a $(\min \{ L, \sqrt{J/\err} \})$-fold advantage.

\FigToy

In practice, we find that learning via OTOCs remains robust even to relatively large amounts of imperfect time-reversal [Fig.~\ref{fig: weak}(d)].
We study this numerically in the ``weak interaction'' learning problem of Fig.~\ref{fig: weak}(b).
As a concrete instance of imperfect time-reversal, we take the spins to be coupled to an extrinsic cavity mode and assume that the spin dynamics are perfectly reversed but the cavity dynamics and spin-cavity coupling $g$ are unreversed.
We find that access to OTOCs substantially improves the classification accuracy even for quite large spin-cavity couplings $g \sim 0.5$, up to half the spin-spin interaction strength.

We can also examine the dependence of learning on read-out errors, namely where one measures a correlator $C$ up to additive error. 
Indeed, we have already incorporated a realistic read-out error $\delta = 3\%$ in our previous numerical studies [Figs.~\ref{fig: probe}(a),~\ref{fig: weak}(a),~\ref{fig: error}(a)].
Intuitively, we expect larger read-out errors to make learning more difficult; however, we have little reason to expect read-out error to change the \emph{relative} advantage of OTOCs compared to TOCs.
We test this numerically by repeating the analysis of Fig.~\ref{fig: weak}(b) for various read-out errors, $\delta$.
For each $\delta$, we compute the minimum link strength $\Jl^*$ that can be learned with $>90 \%$ accuracy [Fig.~\ref{fig: weak}(c)].
For errors $\delta \gtrsim 10^{-3}$, our results agree well with analytic estimates, which predict $(\Jl^*/J)^2 \sim \delta$ for TOCs and $(\Jl^*/J)^2 \sim \delta / L$ for OTOCs.
Intriguingly, for sufficiently small errors $\delta \lesssim 10^{-3}$, the minimum link strength detectable with TOCs \emph{saturates} to a finite value $\Jl^* \sim 0.2$.
Below this value, sample-to-sample fluctuations of the TOC cause the learning task to be difficult regardless of the read-out error.


\section{Provable learning advantage}

We have so far demonstrated the learning power of OTOCs using phenomenological arguments and numerical simulations, owing to the difficulty of obtaining analytic results for ergodic Hamiltonian systems. 
Complementary to these results, we now introduce a binary classification task in which the OTOC is provably efficient.
The task is as follows:
\newline
\newline
\textbf{Disjoint unitary problem:} One is given oracle access to either: ($i$) a fixed, $n$-qubit Haar-random unitary $U$, or ($ii$) a tensor product of two fixed, $n/2$-qubit Haar-random unitaries, $U_1 \otimes U_2$.  The task is to determine which of (i) or (ii) is realized.
\newline

\noindent Qualitatively, this problem resembles the Hamiltonian learning scenarios identified previously. 
First, the feature we seek to learn---the connectivity of the unitary---directly determines how information spreads through the system, as measured by the OTOC.
Second, a Haar-random unitary is inherently ``strongly-interacting'', which causes time-ordered measurements to decay and thus provide little information.

In Fig.~\ref{fig:toy} we show that the disjoint unitary problem can be solved with a constant number (with respect to $n$) of queries to the oracle \emph{and its time-reverse} $U^\dagger$, by measuring an out-of-time-order observable.  Letting $V$ denote the unknown unitary (either $U$ or $U_1 \otimes U_2$), the OTOC is
\begin{equation*}
\textsf{OTOC}(V) = \text{tr}\!\left(\mathds{1}_{\frac{n}{2}} \! \otimes |0\rangle \langle 0|^{\otimes \frac{n}{2}}\!\left\{V^\dagger \sigma_x^1 V |0\rangle \langle 0|^{\otimes n} V^\dagger \sigma_x^1 V\right\}\right).
\end{equation*}
In case ($i$), the OTOC is near zero with probability exponentially close to one~\cite{cotler2022information}.
In case ($ii$), the OTOC is one, since the two subsystems are not coupled by $U_1 \otimes U_2$.
Thus, with probability exponentially close to one, the two cases may be distinguished with a single query to the unknown unitary and its time-reverse.
In contrast, in a companion work~\cite{cotler2022information}, we prove that any time-ordered learning protocol requires an exponential number $\Omega(2^{n/4})$ of queries of the unknown unitary to solve the disjoint unitary problem.
Our proof applies even to adaptive measurement strategies, and leverages novel contemporary techniques from quantum learning theory~\cite{aharonov2021quantum, chen2021exponential, huang2021quantum}.

\section{Discussion}

In this work, we have shown that out-of-time-order measurements can provide powerful advantages for learning the dynamics of quantum systems.
Our results thus highlight the potential gains that can be achieved by quantum experiments if they have sufficient control and coherence to apply time-reversed dynamics.
Extraordinary experimental progress has led to an ever-increasing number of such platforms~\cite{li2017measuring,garttner2017measuring,landsman2019verified,blok2020quantum,mi2021information,sanchez2021emergent,dominguez2021decoherence}, and we envision that learning via OTOCs might find applications across these diverse physical contexts.
Specific future directions include learning long-range cross-talk in quantum processors~\cite{sarovar2020detecting}, and strongly-interacting problems in NMR~\cite{o2021quantum}.

On the theoretical front, our results follow in the footsteps of recent works in quantum learning theory~\cite{aharonov2021quantum,huang2021quantum,chen2021exponential,chen2021hierarchy,cotler2021revisiting} to provide new avenues for quantum advantage. 
Our applications pertain to genuine questions of experimental interest, providing a new bridge between the theoretical tools of quantum learning theory and problems of practical importance in experiments.

\emph{Acknowledgements}---We are grateful to Ryan Babbush, Soonwon Choi, Kostyantyn Kechedzhi, Bryce Kobrin, Lev Ioffe, Vadim Smelyanskiy and Norman Y. Yao for insightful discussions.
The numerical simulations performed in this work used the dynamite Python frontend~\cite{dynamite}, which supports a matrix-free implementation of Krylov subspace methods based on the PETSc and SLEPc packages~\cite{hernandez2005slepc}.
T.S. acknowledges support from the National Science Foundation Graduate Research Fellowship
Program under Grant No. DGE 1752814.  J.C. is supported by a Junior Fellowship from the Harvard Society of Fellows, the Black Hole Initiative, as well as in part by the Department of Energy under grant {DE}-{SC0007870}.

\bibliographystyle{apsrev4-1}
\bibliography{refs_geometry_otocs}

\begin{thebibliography}{68}
\bibitem [{fn4()}]{fn4}%
  \BibitemOpen
  \href@noop {} {\bibinfo  {journal} {To see this explicitly, we specify to the
  case that $A_k, B_l$ are Pauli operators. Consider four possible operations
  on the density matrix $\rho$: the two unitary rotations $\rho \rightarrow
  \frac{\mathbbm{1} \pm iA_k}{\sqrt{2}} \rho \frac{\mathbbm{1} \mp
  iA_k}{\sqrt{2}}$ (where the signs are chosen together) and the two
  projections $\rho \rightarrow \frac{\mathbbm{1} \pm A_k}{2} \rho
  \frac{\mathbbm{1} \pm A_k}{2}$. By taking linear combinations of these four
  possibilities in classical post-processing, we can in effect implement the
  matrix multiplications $\rho \rightarrow A_k \rho$ (and similarly $\rho
  \rightarrow \rho B_l$). Applying the appropriate operation at each time $t_k$
  (and $t'_l$) gives Eq.~(\ref{eq: TOC generic def})}\ }\BibitemShut {NoStop}%
\end{thebibliography}

\pagebreak
\onecolumngrid
\appendix

\section{Details of numerical simulations} \label{sec: numerical}

Here we provide further details on the numerical simulations displayed in Figs.~\ref{fig: probe},~\ref{fig: weak},~\ref{fig: error} of the main text.

\subsection{Correlation functions} \label{sec: correlation functions}

We begin by explicitly writing down the correlation functions used in Figs.~\ref{fig: probe},~\ref{fig: weak},~\ref{fig: error}. 
Throughout, we denote time-evolved operators as $V(t) \equiv U(t,0) V U(t,0)^\dagger$, where the time-evolution unitary is $U(t_2,t_1) = \mathcal{T} \left\{ e^{-i \int_{t_1}^{t_2} dt \, H(t)} \right\}$ and $H(t)$ is the time-dependent stroboscopic Floquet Hamiltonian specified in the main text (unless otherwise stated, in Fig.~\ref{fig: probe ham}).

In the restricted access scenario considered in Figs.~\ref{fig: probe},~\ref{fig: probe ham}, we use the following correlation functions:
\begin{equation} \label{cf probe local}
\begin{split}
C_{\text{TOC}}(t) & = \left\langle V_p(t)  \, V_p(0) \right\rangle \\
C_{\text{TOC}}(x,t) & = \left\langle V_p(t) \, W_x(t/2) \, V_p(0) \, W_x(t/2) \right\rangle \\
C_{\text{OTOC}}(x,t) & = \left\langle V_p(t) \, W_x(0) \, V_p(t) \, W_x(0) \right\rangle
\end{split}
\end{equation}
where $p$ denotes the probe qubit, and $\langle \cdot \rangle \equiv 2^{-L} \tr( \cdot )$ is an infinite temperature average.
Each of these correlation functions can be measured using state preparation and read-out on the probe qubit, combined with time-evolution and a single local unitary operation on the larger system.
(In the case of the auto-correlation function $C_{\text{TOC}}(t)$ no local unitary operation is needed, $W_x = \mathbbm{1}$.)

In principle, we envision allowing $V, W$ to run over all local operators in the system.
For instance, they could run over all $4^w \binom{N}{w}$ Pauli operators of weight $\leq w$, where $w \sim O(1)$.
This is naturally achieved by randomized measurement strategies such as shadow tomography with local Clifford unitaries and $O(3^w)$ measurements~\cite{cotler2020quantum,huang2020predicting}.
In practice, we must restrict $V, W$ to a few possible values in numerical simulations.
Specifically, we take $V = W \in \{\sigma_x, \sigma_z\}$ for TOCs, and $V = W \in \{\sigma_z\}$ for OTOCs.
The OTOC is observed to be relatively insensitive to basis of $V$ and $W$, hence our choice to restrict to a single operator, $\sigma_z$ (further, we note that adding $\sigma_x$ OTOCs could only improve the relative advantage of OTOCs compared to TOCs).
More broadly, we do not expect that adding additional pairs of $\{ V, W \}$ will change the qualitative behavior of learning via TOCs and OTOCs.
Specifically, we have seen that the learning advantage of OTOCs arises from their ability to detect highly non-local correlations in the system (i.e.~large-weight components of the time-evolved operator $V_p(t)$, see Appendix~\ref{sec: Fisher restricted} for more detailed phenomenological estimates).
These correlations are not detectable by any time-ordered correlator involving only few-body operators; indeed, in ergodic systems we generically expect that they are not efficiently detectable by \emph{any} time-ordered measurement.

For Figs.~\ref{fig: weak},~\ref{fig: error} of the main text, we utilize two-point correlation functions between pairs of local operators:
\begin{equation}\label{eq: corr fns fig 3bd}
\begin{split}
C_{\text{TOC}} & = \left\langle V_x(t) \, W_{x'}(0) \right\rangle, \\
C_{\text{OTOC}} & = \left\langle V_x(t) \, W_{x'}(0) \, V_x(t) \, W_{x'}(0) \right\rangle.
\end{split}
\end{equation}
We again take $V = W \in \{\sigma_x, \sigma_z\}$ for TOCs and $V = W \in \{\sigma_z\}$ for OTOCs.
We allow $x, x'$ to span all qubits within a distance 2 of the link---this consists of 6 possible values for each of $x, x'$, corresponding to distances 0, 1, and 2 to both the left and right of the link.
In principle, we would like $x, x'$ to run over the entire lattice; however, in practice we observe that correlation functions involving qubits distant from the link provide little information, and so can be safely neglected.

We now briefly comment on our numerical methods for computing the above correlation functions and the Fisher information [Figs.~\ref{fig: probe}(b),~\ref{fig: weak}(b)].
We compute the infinite temperature average in the correlation functions by sampling over Haar-random initial states $\ket{\psi}$.
To motivate this, we can insert a resolution of the identity, $\mathbbm{1} = \frac{1}{2^L} \sum_\psi \dyad{\psi}$ into the correlation functions Eq.~(\ref{eq: corr fns fig 3bd}) to obtain:
\begin{equation} \label{eq: corr fns Haar random}
\begin{split}
C_{\text{TOC}} & =  \frac{1}{2^L} \sum_\psi \bra{\psi} V_i(t) \, W_j(0) \ket{\psi}\,,\\
\end{split}
\end{equation}
and similarly for the OTOC.
In numerics, we approximate this sum by sampling a finite number $N_\psi$ of states $|\psi\rangle$ drawn from the Haar distribution; errors in this approximation will scale as $\sim 1/\sqrt{N_\psi 2^L}$.

\FigExtrapolate

In the learning problems considered in the main text [Fig.~\ref{fig: probe}(a),~\ref{fig: weak}(a),~\ref{fig: error}], we take $N_\psi = 25, 25, 10, 1, 1$ for system sizes $L= 6,8,10,12,14$, respectively.
In contrast, when estimating the Fisher information [Fig.~\ref{fig: probe}(b),~\ref{fig: weak}(b)], we perform a large-$N_\psi$ extrapolation to improve precision.
This is required in order to establish the asymptotic scaling of the Fisher information at large $d$ [Fig.~\ref{fig: probe}(b)] and small $J_\ell$ [Fig.~\ref{fig: weak}(b)].
Specifically, we compute the estimated correlation function $C_{N_\psi}$ averaged over $N_\psi = 1, \ldots, 25$ Haar-random initial states, as well as the resultant Fisher information $\text{max FI}(N_\psi) \equiv | \partial C_{N_\psi} / \partial J |^2$, maximized over all relevant correlation functions.
We then perform a linear fit $\text{max FI}(N_\psi) = \text{max FI}(\infty) + \frac{A}{N_\psi}$, where $\text{max FI}(\infty)$ and $A$ are fitting parameters.
Finally, the fitting parameter $\text{max FI}(\infty)$ represents our estimation of the Fisher information at $N_\psi \rightarrow \infty$, which we plot in Figs.~\ref{fig: probe}(b),~\ref{fig: weak}(b).
We illustrate this procedure in Fig.~\ref{fig: extrapolate}, using the data for Fig.~\ref{fig: probe}(b).
On the left of Fig.~\ref{fig: extrapolate}, we plot the maximum Fisher information, $\text{max FI}(N_\psi)$, for each $N_\psi$, as a function of the distance $d$.
We observe that in regions where $C$ is relatively large (i.e. small $d$), the estimates are quite accurate even for $N_\psi = 1$, while in regions where $C$ is small (i.e.~large $d$) the Fisher information becomes successively smaller as the number of sampled states $N_\psi$ increases.
On the right of Fig.~\ref{fig: extrapolate}, we re-plot the Fisher information for each $d$ as a function of $N_\psi$.
Solid lines represent the results of the linear fit, which we observe to fit the $N_\psi$-dependence of the data quite well.
The extrapolated Fisher information [as displayed in Fig.~\ref{fig: probe}(b)] is shown in Fig.~\ref{fig: extrapolate} as a dashed line.


\subsection{Imperfect time-reversal via cavity mode}

In Fig.~\ref{fig: error}(a), we benchmark the effects of decoherence on learning by coupling the spin system to a single cavity mode.
Our motivation for studying this model is two-fold.
First, in ergodic many-body systems the effect of local errors on OTOCs is expected to be independent of the precise microscopic form of the error~\cite{schuster2022time}.
We therefore expect the spin-cavity system to display similar OTOC physics to more generic local error models.
Second, for $L=10$ spins the spin-cavity system can be exactly simulated in a Hilbert space of size $2^L \times L$ (we assume the cavity initially has zero occupation number; since the sum of spin magnetization and the cavity occupation is conserved, the cavity occupation is upper bounded by $L$).
This is substantially smaller than the requirements to exactly simulate a mixed state quantum system, $~2^{2L}$.

More specifically, the spin-cavity Hamiltonian is as follows.
We modify the Floquet time-evolution described in the main text to alternate between the following two Hamiltonians:
\begin{equation}
\begin{split}
H_1 & = \pm H_f + g \sum_i \left( a^\dagger \sigma_i^- + a \sigma_i^+ \right) + \omega a^\dagger a \\
H_2 & = \pm H_c + g \sum_i \left( a^\dagger \sigma_i^- + a \sigma_i^+ \right) + \omega a^\dagger a
\end{split}
\end{equation}
where $H_f, H_c$ are the field and coupling Hamiltonians written in the main text, $a, a^\dagger$ are lowering/raising operators for a bosonic cavity mode, $g = \{ 0.0, 0.25, 0.5\}$ is the spin-cavity interaction strength, and $\omega = 1.7$ is the cavity frequency.
Here, the $\pm$ denote values during forwards/backwards time-evolution; note that we do \emph{not} reverse the spin-cavity interaction or the cavity frequency during backwards time-evolution.

\subsection{Learning model} \label{sec: learning details}

We now detail the machine learning techniques used in Figs.~\ref{fig: probe}(a),~\ref{fig: weak}(a),~\ref{fig: error} of the main text, and Fig.~\ref{fig: threeway}(b) of the Appendix.
Throughout, read-out error is mimicked by adding a random Gaussian variable with mean zero and standard deviation $\delta$ to the exact correlation functions.

We begin with Fig.~\ref{fig: probe}(a).
Our goal is to predict the value of $d$ [which specifies the geometry of the spin system, see Fig.~\ref{fig: probe}(a)] from the correlation functions of the system, Eq.~(\ref{cf probe local}).
To do so, we train a learning model on 3000 randomly drawn disorder realizations of the Hamiltonian, consisting of 300 realizations each for $d = 0,1,\ldots,9$.
We test model performance on 2000 additional disorder realizations, again consisting of 200 realizations each for $d = 0,1,\ldots,9$.
For each disorder realization, the input to our learning model consists of the correlation functions Eq.~(\ref{cf probe local}), evaluated at $x = 2,\ldots,L$ and $30$ evenly spaced times between $0$ and $12$.
We apply Gaussian distributed read-out error $\delta = 3\%$ to each correlation function.
We repeat this procedure, as well as the model training and evaluation that follows, first using only TOCs as input to the learning algorithm, and second using both TOCs and OTOCs.

Next, we input these correlation functions into a support vector regression (SVR) model with radial basis functions~\cite{sklearn_api}.
The radial SVR contains two hyperparameters: $C$, the regularization parameter, and $\gamma$, which controls the width of the radial basis functions.
We choose $C$ and $\gamma$ by performing five-fold cross-validation over the sets $C = \{10,30,100\}$, $\gamma =\{ 0.03,0.06,0.1,0.3,1,3,6,10\}$.
We obtain $C = 10, \gamma = 1$ for learning via TOCs, and $C = 10, \gamma = 0.03$ for learning via both TOCs and OTOCs.
In the identical learning task for Hamiltonian evolution [Fig.~\ref{fig: probe ham}(a)], we obtain $C = 10, \gamma = 1$ for learning via TOCs, and $C = 100, \gamma = 0.03$ for learning via both TOCs and OTOCs.

We now turn to Fig.~\ref{fig: weak}(a) and~\ref{fig: error}.
Our goal is to perform binary classification using the correlation functions Eq.~(\ref{eq: corr fns fig 3bd}) to distinguish whether the link interaction strength is zero or nonzero.
To do so, we simulate the correlation functions of 300 randomly drawn disorder realizations of the Hamiltonian for each link strength, $J_\ell = \{0, 0.01,0.017,0.03,0.06,0.1,0.17,0.3,0.6,1.0 \}$.
For each nonzero $J_\ell$, we train a learning model to perform binary classification between link strength $0$ and $J_\ell$.
We test model performance on 400 additional disorder realizations, again consisting of 200 realizations each for link strength $0$ and $J_\ell$.

The first step of our learning model is to prune the correlation functions used as input.
We do so by estimating the mutual information between each individual correlation function and the link interaction strength, and selecting the $K$ correlation functions with the highest mutual information. Here $K$ is a hyperparameter that will ultimately be chosen via cross-validation.
To estimate the mutual information, we fit the distribution of correlation functions values over disorder realizations to a Gaussian for each link strength, and compute the Jensen-Shannon divergence between the Gaussian distributions.
The Jensen-Shannon divergence is equal to the desired mutual information~\cite{lin1991divergence}.

As before, we input the selected correlation functions into a support vector machine (SVM) with radial basis functions~\cite{sklearn_api}.
We now have three hyperparameters: the SVM hyperparameters $C$ and $\gamma$ and the number of selected correlation functions $K$.
We choose $C$, $\gamma$, and $K$ by performing five-fold cross-validation over each value $C = \{0.1,1,10,100,1000\}$, $\gamma =\{ 0.1, 0.3, 1,3,10,30\}$.
We obtain Figs.~\ref{fig: error}(b), by repeating this procedure for various simulated read-out errors, $\delta = \{0.0001, 0.0003, 0.001, 0.003, 0.01, 0.03, 0.1\}$.
At each read-out error, we perform a linear interpolation of the classification accuracy as a function of $J_\ell$ [as shown in Fig.~\ref{fig: weak}(a) for $\delta = 0.03$].
The minimum detectable link strength $J^*_\ell$ occurs at the intersection of this interpolation with a horizontal line (not depicted) corresponding to a classification accuracy of $90\%$.

Finally, we turn to Fig.~\ref{fig: threeway}(b) in  Appendix~\ref{sec: additional numerics}.
In the probe qubit scenario, training and testing are performed on $300$ and $200$ samples respectively for each geometry and each value of $d$.
In the global state preparation and read-out scenario, we instead use $60$ and $40$ samples respectively for each geometry.
Our learning model consists of a support vector machine with hyperparameters chosen via 4-fold cross-validation from the sets $C = \{0.1,1,10\}$, $\gamma =\{ 0.3, 3, 30\}$.
As in our previous learning tasks, we apply a read-out error $\delta = 3\%$ to each correlation function before use in learning.

\section{Phenomenological estimates} \label{sec: Fisher restricted}

In this section, we provide more detailed reasoning behind the phenomenological estimates of the Fisher information presented in the main text.
We begin with the Fisher information under unitary dynamics with restricted access and then turn to the effects of imperfect time-reversal and decoherence.

\subsection{Fisher information in restricted access scenario}

At sufficiently large times and distances, we expect the profile of correlation functions in ergodic many-body systems to be described by only a few phenomenological parameters.
For instance, in one-dimensional systems the out-of-time-order correlator is predicted to take the following functional form~\cite{aleiner2016microscopic,nahum2017quantum},
\begin{equation} \label{eq: OTOC f form}
    C_{\text{OTOC}}(x,t) \approx f\left( \frac{x/v_B - t}{A\sqrt{t}} \right)
\end{equation}
where the phenomenological parameters $v_B$ and $A$ describe the butterfly velocity and the width of the OTOC wavefront, respectively. Here $f$ is a compactly supported bump function which interpolates between zero and one and then zero again within an $O(1)$-sized region about the origin.
Meanwhile, in systems with a local conservation law, we expect time-ordered correlators to be dominated by diffusion of the conserved quantity.
This leads to the following profile for the auto-correlation function,
\begin{equation} \label{eq: TOC diffusion}
    C_{\text{TOC}}(t) \sim \frac{1}{\sqrt{D t}},
\end{equation}
where $D$ is a diffusion constant.
In the absence of conserved quantities, one expects time-ordered correlation functions to instead decay exponentially in time,
\begin{equation} \label{eq: exp decay}
    C_{\text{TOC}}(t) \sim \exp( - \gamma t ),
\end{equation}
parameterized by a decay rate $\gamma$.

To obtain the Fisher information, $\text{FI}(J | C) = \left| \partial C / \partial J \right|^2$, we must compute the derivative of the correlation functions with respect to a local coupling strength, $J_y$.
To do so while leveraging the above phenomenological predictions, we must first recognize that the phenomenological parameters are themselves dependent on the local coupling strengths of the system, e.g. $v_B \rightarrow v_B(\{ J_y \})$.
We expand on this in further detail for each case below.
The resultant scaling of the Fisher information in various physical regimes is summarized in Table~\ref{table restricted}.

\begin{table}[t]
\vspace{-2.3mm}
\caption{Maximum Fisher information in restricted access scenarios} 
\centering 
\begin{tabular}{c c c} 
\hline\hline 
Correlation function & \,\,\,\,\,\, Probe qubit with \,\,\,\,\,\, & Probe qubit with
\\ [0.0ex]
 & \,\,\,\,\,\, local unitary control \,\,\,\,\,\, & global unitary control
\\ [0.5ex]
\hline 
&  & \\[-2ex]
TOC without & $O(\exp(-d))$, & $O(\exp(-d))$,\\[-0.ex]
conserved quantity & [Fig.~\ref{fig: probe}(b)] &   \\[-1ex]
&  & \\[-0.1ex]
TOC with &  $O(1/d^4)$ & $O(1/d^4)$ \\[-0.ex]
conserved quantity &  & \\[-1ex]
&  & \\[-0.1ex]
OTOC  &  $O(1/d)$, & $O(1/d^2)$, \\[-0.ex]
& [Fig.~\ref{fig: probe}(b)] &  \\[-1ex]
&  & \\[-1.5ex]
\hline 
\vspace{-3.3mm}
\end{tabular}
\label{table restricted}
\caption{Phenomenological estimates of the scaling of the Fisher information in the restricted access scenario, for learning an interaction that lies a distance $d$ from the probe qubit.}
\end{table}

\subsubsection{Fisher information of OTOCs}

We begin with the Fisher information of OTOCs.
Our treatment is broken into two parts, corresponding to the scenarios where the experimenter has either local or global unitary control over the larger system.
The former scenario is simulated numerically in Fig.~\ref{fig: probe} of the main text.

\emph{Local unitary control.}---We consider local OTOCs [Eq.~(\ref{cf probe local})] and are interested in the dependence of the OTOC on the local coupling strengths, $\{ J_y \}$.
To approach this, we will assume that OTOC takes the same functional form as in Eq.~(\ref{eq: OTOC f form}),
\begin{equation}
C_{\text{OTOC}}(x,t) = \left\langle V_p(t) \, W_x(0) \, V_p(t) \, W_x(0) \right\rangle \approx f\left( \frac{x / v_B(x) - t}{A\sqrt{t}} \right),
\end{equation}
but now with a \emph{position-dependent} butterfly velocity, $v_B(x)$.
Specifically, we assume that the effective butterfly velocity at time $t$ receives contributions from all couplings that have been visited thus far, i.e.~all $J_y$ with $y \lesssim x$.
Since the time to traverse a single coupling is proportional to the inverse coupling strength $1/J_y$, we expect the time to traverse all couplings up to a distance $x$ to be proportional to the sum $\sum_{y=0}^x 1/J_y$.
Equating this time to the distance divided by the effective butterfly velocity $x / v_B(x)$, we have:
\begin{equation}
    v_B(x) \approx \left[ \frac{1}{x} \sum_{x=0}^{x} \frac{1}{J_x} \right]^{-1}.
\end{equation}
If each coupling strength is drawn independently from some disorder realization, then at large times the butterfly velocity will be close to its typical value, $v_B = \overline{1/J}$.
 
We can now compute derivatives of the correlation function with respect to a given coupling strength via the chain rule.
The derivative of the butterfly velocity is
\begin{equation}
    \partial_{J_d} v_B(x) \approx \frac{v_B(x)^2}{J_d^2 x} \cdot \delta_{d \leq x},
\end{equation}
which yields the following for the OTOC:
\begin{equation} \label{OTOC derivative}
\begin{split}
    \partial_{J_d} C_{\text{OTOC}}(x,t) & \approx  - \frac{x}{A\sqrt{t} v_B(x)^2} \cdot f'\left( \frac{x/v_B(x) - t}{A\sqrt{t}} \right) \cdot \partial_{J_d} v_B(x)  \\
    & \approx  - \frac{1}{A J_d^2\sqrt{t}} \cdot \delta_{d \leq x} \cdot f'\left( \frac{x/v_B(x) - t}{A\sqrt{t}} \right). \\
\end{split}
\end{equation}

There are two parameters of the local OTOC chosen by a potential experimentalist: the position $x$ of the local perturbation, and the evolution time $t$.
We are interested in the maximum Fisher information given an optimal choice of $x$ and $t$.
Observing Eq.~(\ref{OTOC derivative}), we see that the derivative $f'$ is maximized by the choice $x = v_B t$, while the delta function then sets $v_B t = d$.
Plugging these values in, we find the Fisher information
\begin{equation}
    \max_C \text{FI}(J | C_{\text{OTOC}}) \approx \left| \frac{f'(0)}{A J_y^2 \sqrt{v_B d}} \right|^2 \sim \frac{1}{d}.
\end{equation}

\emph{Global unitary control.}---We now turn to an alternate experimental scenario, where one has only global unitary control over the larger system.
In this scenario, the natural generalization of the correlation functions Eq.~(\ref{cf probe local}) is the following:
\begin{equation} \label{cf probe global}
\begin{split}
C_{\text{TOC}} & = \left\langle V_p(t)  \, V_p(0) \right\rangle \\
C_{\text{TOC}} & = \left\langle V_p(t) \, e^{i \phi \sum_x W_x(t/2)} \, V_p(0) \, e^{-i \phi \sum_x W_x(t/2)} \right\rangle \\
C_{\text{OTOC}} & = \left\langle V_p(t) \, e^{i \phi \sum_x W_x(0)} \, V_p(t) \, e^{-i \phi \sum_x W_x(0)}. \right\rangle
\end{split}
\end{equation}
Here we replace the local unitary operations of Eq.~(\ref{cf probe local}) with global spin rotations, $e^{i \phi \sum_x W_x}$, by an angle $\phi$ (here, $W_x$ is a local Hermitian operator on qubit $x$).
%

%
We expect the behavior of the OTOC under global control to be governed by the ``size'' of time-evolved operators~\cite{sanchez2019emergent,sanchez2021emergent,dominguez2021decoherence}.
The size corresponds to the average of local OTOCs over all qubits in the system~\cite{qi2019quantum}.
In one-dimensional ergodic systems the size grows linearly $\sim v_B t$, which yields the following phenomenological expectation for the global OTOC~\cite{dominguez2021decoherence}:
\begin{equation}
    C^{\text{glob}}_{\text{OTOC}} = \exp( - \phi^2 v_B(t) t ).
\end{equation}
Here we have made the butterfly velocity time-dependent to capture its dependence on the local coupling strengths,
\begin{equation}
    v_B(t) \approx \left[ \frac{1}{v_B t} \sum_{y=0}^{v_B t} \frac{1}{J_y} \right]^{-1},
\end{equation}
where $v_B = \overline{1/J}$ is the typical butterfly velocity.

Taking the derivative of the OTOC via the chain rule, we have:
\begin{equation}
\begin{split}
    \partial_{J_d} C^{\text{glob}}_{\text{OTOC}} & \approx  -\phi^2  t \cdot \partial_{J_d} v_B(t) \cdot \exp(-\phi^2 v_B(t) t) \\
    & =  - \phi^2  t \cdot \frac{v_B(t)^2}{v_B J_d^2 t} \cdot \delta_{d \leq v_B(t) t} \cdot \exp(-\phi^2 v_B(t) t) \\
    & \approx  - \phi^2 \cdot \frac{v_B}{J_d^2} \cdot \delta_{d \leq v_B t} \cdot \exp(-\phi^2 v_B t)\,. \\
\end{split}
\end{equation}
We would like to maximize the Fisher information over the parameters $(\phi, t)$.
This entails taking the time $t$ to be as early as allowed by the delta function, $t \approx d/v_B$, in order to minimize the exponential.
The correlator is then maximized by choosing $\phi$ such that $\phi^2 v_B t \sim 1$.
This gives a Fisher information:
\begin{equation} \label{fisher restricted OTOC}
    \max_C \text{FI}(J | C_{\text{OTOC}}) \approx \left| \frac{v_B e^{-1}}{J_y^2 d} \right|^2 \sim \frac{1}{d^2},
\end{equation}
which decays algebraically, with an additional factor of $d$ compared to the local unitary control scenario.

Before moving on, we briefly summarize the intuition behind the two above estimates.
In both cases, an $O(1)$ perturbation in a local coupling strength produces an $O(1)$ shift in the location of the OTOC wavefront.
With local control, this shift produces an $O(1/\sqrt{d})$ change in the OTOC, since the OTOC wavefront is spread across a width $\sim \! \sqrt{d}$ by the time it reaches the coupling.
With global control, this produces an $O(1/d)$ change in the OTOC, since the global OTOC depends on the average of $\sim \!  d$ individual coupling strengths.
Since the Fisher information involves the square of the OTOC derivative, these lead to an $O(1/d)$ and $O(1/d^2)$ Fisher information, respectively.

\subsubsection{Fisher information of TOCs in absence of conserved quantities}

We now turn to a simpler case, the Fisher information of time-ordered correlators in the absence of conserved quantities [Fig.~\ref{fig: probe}(b)].
Under ergodic dynamics, we expect such correlation functions to decay exponentially in time at sufficiently large times, see Eq.~(\ref{eq: exp decay}).
Now, consider the derivative of the correlation function with respect to a local coupling strength at a distance $d$ away from the probe qubit.
By causality, this derivative can only be non-zero after a time $t \gtrsim d/v_B$.
However, at such times the magnitude of the correlation function has already decayed by a factor of $e^{-\gamma t}$.
This suggests that the Fisher information will decay exponentially in the distance $d$,
\begin{equation}
    \max_C \text{FI}(J | C_{\text{TOC}}) \lesssim \exp( - 2 \gamma x / v_B ),
\end{equation}
as observed numerically in Fig.~\ref{fig: probe}(b).

\subsubsection{Fisher information of TOCs in presence of conserved quantities} \label{sec: conserved}

The scaling of the Fisher information for TOCs is modified in the presence of a conserved quantities.
In this case, one expects the TOC at sufficiently large times to be dominated by slow diffusive dynamics of the conserved quantity.
This lies in contrast to the exponential decay expected in the absence of conserved quantities. 

To study this, we begin with the auto-correlation function [first line of Eq.~(\ref{cf probe local})].
Recall that the auto-correlation function can be measured with access solely to the probe qubit, and is thus accessible in both the local and global control scenarios.
%
Similar to the case of OTOCs, we will assume that the dependence of the correlation function on the local coupling strengths is captured by replacing the diffusion constant, $D$, with a \emph{time-dependent} value,
\begin{equation}
    C_{\text{TOC}}(t) = \frac{1}{\sqrt{D(t) t}}.
\end{equation}
Following the logic of the previous section, we assume the effective diffusion constant takes the form~\cite{revathi1993effective},
\begin{equation}
    D(t) \sim \left[ \frac{1}{\sqrt{D t}} \sum_{y=0}^{\sqrt{D t}} \frac{1}{J_y} \right]^{-1},
\end{equation}
where $D = \overline{1/J}$ is the diffusion constant's typical value.
%
%
Differentiating with respect to the local coupling strength gives
\begin{equation}
    \partial_{J_d} D(t) \sim \frac{D(t)^2}{J_d^2 \sqrt{D t}} \cdot \delta_{d < \sqrt{Dt}}\,.
\end{equation}

Computing the derivative of the auto-correlation function, we have:
\begin{equation}
\begin{split}
    \partial_{J_d} C_{\text{TOC}}(t) & = -\frac{1}{2 D(t)^{3/2} t^{1/2}} \cdot \partial_{J_d} D(t) \\
    & = - \frac{D(t)^{1/2}}{2 J_d^2 t \sqrt{D}} \cdot \delta_{d < \sqrt{Dt}} \\
    & \approx - \frac{1}{2 J_d^2 t} \cdot \delta_{d < \sqrt{Dt}}\,. \\
\end{split}
\end{equation}
The magnitude of the derivative is maximized by taking the minimum possible time, $t \approx d^2 / D$, which yields a Fisher information,
\begin{equation}
    \max_C \text{FI}(J | C_{\text{TOC}}) \approx \left| \frac{D}{2 J_x^2 d^2} \right|^2 \sim \frac{1}{d^4}.
\end{equation}
This can be understood intuitively as follows.
For the auto-correlation function to be sensitive to the coupling strength $J_d$, the conserved quantity must have spread to at least distance $d$.
At such a distance the magnitude of the auto-correlation function is $O(1/d)$, since the conserved quantity has spread over $\sim d$ sites.
In addition, the derivative with respect to an individual coupling strength is suppressed by an additional factor $O(1/d)$, since the auto-correlator depends only on the average (inverse) coupling strength over $\sim d$ sites.
Combining these two factors and squaring leads to an $O(1/d^4)$ Fisher information.

We now turn to the remaining time-ordered correlation functions in Eqs.~(\ref{cf probe local},~\ref{cf probe global}), which require either local or global unitary control over the non-probe qubits.
We will find that such correlators provide no scaling advantage beyond the auto-correlator.

We first consider the case of local unitary control [Eq.~(\ref{cf probe local})].
Physically, these correlation functions correspond to preparing an amount of the conserved quantity (e.g.~a spin polarization) at the probe qubit, letting it diffuse for a time $t/2$, flipping the spin polarization at a qubit $x$, and measuring the polarization at the probe qubit after an additional time $t/2$.
We thus expect the TOC behave as follows,
\begin{equation}
    C_{\text{TOC}}(x,t) = \left\langle V_p(t) \, W_x(0) \, V_p(t) \, W_x(0) \right\rangle \approx q(0,t) - 2 q(x,t/2) \cdot q(x,t/2),
\end{equation}
where $q(x,t) \approx (2\pi D(t) t)^{-1/2} \exp( - x^2 / ( 2 D(t) t ) )$ is the propagator of the conserved quantity from position $0$ to position $x$ (or vice versa).
The first term is equal to the auto-correlation function.
The second term arises from the spin flip at position $x$ and time $t/2$.
The spin flip effectively inserts a negative polarization $-2q(x,t/2)$ on the qubit $x$, which propagates back to the probe qubit with amplitude $q(x,t/2)$.

The derivative of the second term is as follows,
\begin{equation}
\begin{split}
    \partial_{J_d} \left[ q(x,t/2)^2 \right] 
    & = \partial_{J_d} \left[ \frac{1}{\pi D(t) t} \exp( - \frac{2 x^2}{ D(t) t } ) \right]  \\
    & \approx \partial_{J_d} \left[ \frac{-1}{\pi J_y^2 D^{1/2} t^{3/2}} \exp( - \frac{2 x^2}{ D(t) t } ) + \frac{2 x^2}{\pi J_y^2 D^{3/2} t^{5/2}} \exp( - \frac{2 x^2}{ D(t) t } ) \right] \cdot \delta_{d < \sqrt{Dt}}\,. \\
\end{split}
\end{equation}
The magnitude of the derivative is maximized at $D t \sim x^2, x \sim d$, and is of order $O(1/d^3)$.
This is subleading compared to the auto-correlation function, of order $O(1/d^2)$, and thus does not affect the asymptotic scaling of the Fisher information with $d$.

The case of a global control [Eq.~(\ref{cf probe global})] is even simpler.
A global spin rotation about the $x$-axis by an angle $\phi$ multiplies the conserved quantity at each site by a factor of $\cos(\phi)$.
Here we assume that the $x$- and $y$-components of spin that are generated by the rotation quickly decay in time if they are not conserved by the ergodic dynamics.
The resulting correlation function is then given by the auto-correlation multiplied by $\cos(\phi)$.
Again, this provides no scaling advantage in the Fisher information.

\subsection{Effect of imperfect time-reversal and decoherence on Fisher information}

We now incorporate imperfect time-reversal dynamics into our estimates of the Fisher information of OTOCs.
Previous works have been found that a wide range of experimental errors (e.g. extrinsic decoherence, coherent errors in time-reversal) have a similar effect on OTOC measurements, as long as the relevant errors are local and the dynamics are ergodic~\cite{swingle2018resilience,mi2021information,sanchez2021emergent,dominguez2021decoherence,schuster2022time}.

Specifically, in one-dimensional systems, one expects that the OTOC under open-system dynamics, $\tilde{C}_{\text{OTOC}}$, is equal to the same OTOC under unitary dynamics, $C_{\text{OTOC}}$, multiplied by an overall Gaussian decay in time~\cite{schuster2022time}:
\begin{equation} \label{peaked size}
\tilde{C}_{\text{OTOC}} \approx \exp( - a \varepsilon v_B t^2 ) \times C_{\text{OTOC}}\,.
\end{equation}
Here $\varepsilon$ is an effective local error rate, $v_B$ is the butterfly velocity, and $a$ is an order one constant.
The argument of the above exponential is proportional the volume of the time-evolved operator's light cone.
Intuitively, Eq.~(\ref{peaked size}) states that each error in the causal past of an operator contributes a roughly equal amount to the decay of the OTOC.
We note that in finite-size systems we do not expect Eq.~(\ref{peaked size}) to precisely hold, however, corrections are expected to be suppressed by $\sim \! \varepsilon / J$ where $J$ is the local interaction strength~\cite{schuster2022time}, so we neglect them here.

Substituting Eq.~(\ref{peaked size}) into our estimate for the Fisher information [Eq.~(\ref{fisher restricted OTOC})] and setting $v_B t \approx d$, we find:
\begin{equation}
\max_C \, \text{FI}(J_d | \tilde{C}_{\text{OTOC}}) \sim \frac{1}{d^2} \exp(- a \varepsilon d^2 /v_B )\,.
\end{equation}
Meanwhile, we assume that the Fisher information with respect to TOCs is comparatively unaffected by error, and thus once again follows a linear exponential decay in $d$:
\begin{equation}
\max_C \, \text{FI}(J_d | \tilde{C}_{\text{TOC}}) \sim \exp(- \gamma d)\,.
\end{equation}
Setting the two exponentials to be equal, $\max_C \, \text{FI}(J_d | \tilde{C}_{\text{OTOC}}) \sim \max_C \, \text{FI}(J_d | \tilde{C}_{\text{TOC}})$, we find that the OTOC continues to provide an advantage over the TOC  up to
\begin{equation}
d \lesssim \frac{\gamma v_B}{\varepsilon}\,,
\end{equation}
as quoted in the main text.

We now apply the same analysis to our second learning regime.
Let us set $v_B \sim J$ for consistency with the main text.  The Fisher information of an OTOC between operators on either side of the link with respect to the link interaction strength is now modified to
\begin{equation}
\max_C \, \text{FI}(J_\ell | \tilde{C}_{\text{OTOC}}) \sim \frac{J_\ell^4 t^2}{J^2} \exp(- a \varepsilon J t^2), \,\,\,\,\,\,\, J t \lesssim L\,.
\end{equation}
The maximum of the Fisher information as a function of time now occurs at
\begin{equation}
t^* \sim \min \left\{ \sqrt{\frac{1}{\varepsilon J}}, \frac{L}{J} \right\},
\end{equation}
with value
\begin{equation} \label{eq: FI OTOC error link}
\max_C \, \text{FI}(J_\ell | \tilde{C}_{\text{OTOC}}) \sim \min \left\{ \frac{J_\ell^4}{\varepsilon J^3} , \frac{J_\ell^4 L^2}{J^4} \right\},
\end{equation}
which differs from the unitary OTOC for sufficiently high error rates.
Again, we assume that the Fisher information of TOC is not affected by error to leading order.
Taking the square root of Eq.~(\ref{eq: FI OTOC error link}), we thus find the $L$-fold advantage of the OTOC is replaced by a $\sqrt{J/\varepsilon}$-fold advantage at error rates $\sqrt{\varepsilon/ J} \gtrsim 1/L$, as quoted in the main text.

\section{Additional numerics}\label{sec: additional numerics}

\FigHamProbe

In this section we provide numerical results in two additional learning scenarios.
We begin by repeating the simulations leading to Fig.~\ref{fig: probe} for time-independent Hamiltonian evolution instead of Floquet evolution.
We then discuss learning in the restricted access scenario with global unitary control, in contrast to local unitary control as considered in the main text).

\subsection{Learning with time-independent Hamiltonian evolution}

In the main text numerical simulations, we utilized Floquet time-evolution in which the spin interactions and local fields were applied in a stroboscopic fashion.
Our motivations for using Floquet time-evolution instead of Hamiltonian time-evolution were three-fold: First, Floquet dynamics are prevalent in a variety of quantum systems that one might wish to learn, e.g.~in digital quantum simulators, and NMR or solid-state defect setups with optical driving.
Second, the Floquet dynamics considered are moderately faster to simulate via Krylov subspace methods than Hamiltonian dynamics, since the Hamiltonian of the former contains fewer terms at a given instant in time.
Third, we do not expect the behavior of learning via TOCs or OTOCs under the two dynamics to qualitatively differ at moderate times and distances (although at large distances they may, see Appendix~\ref{sec: Fisher restricted}).

Here, we check the latter assumption by repeating the numerical analysis of Fig.~\ref{fig: probe} using time-independent Hamiltonian dynamics.
As shown in Fig.~\ref{fig: probe ham}(a), we find that the results of the learning task of Fig.~\ref{fig: probe}(a) behave quite similarly for Hamiltonian and Floquet dynamics.
In particular, access to OTOCs continues to enable substantially more accurate predictions for the crossing distance $d$ for all $d \gtrsim 3$.
In Fig.~\ref{fig: probe ham}(b), we turn to the behavior of the Fisher information as a function of a coupling's distance from the probe qubit.
Unfortunately, we are not able to discern the $\sim \! 1/d^4$ scaling predicted in Appendix~\ref{sec: Fisher restricted} in our finite-size numerics.
Instead, the Fisher information behaves qualitatively similar to that of Floquet dynamics [Fig.~\ref{fig: probe}(b)].
We anticipate that at sufficiently large distances the Fisher information of Hamiltonian dynamics will indeed asymptote to the expected power law decay. However, at such distances the Fisher information will likely already be too small to be useful for most practical purposes.

\subsection{Learning under restricted access with global unitary control}

\begin{figure}
\centering
\includegraphics[width=0.7\textwidth]{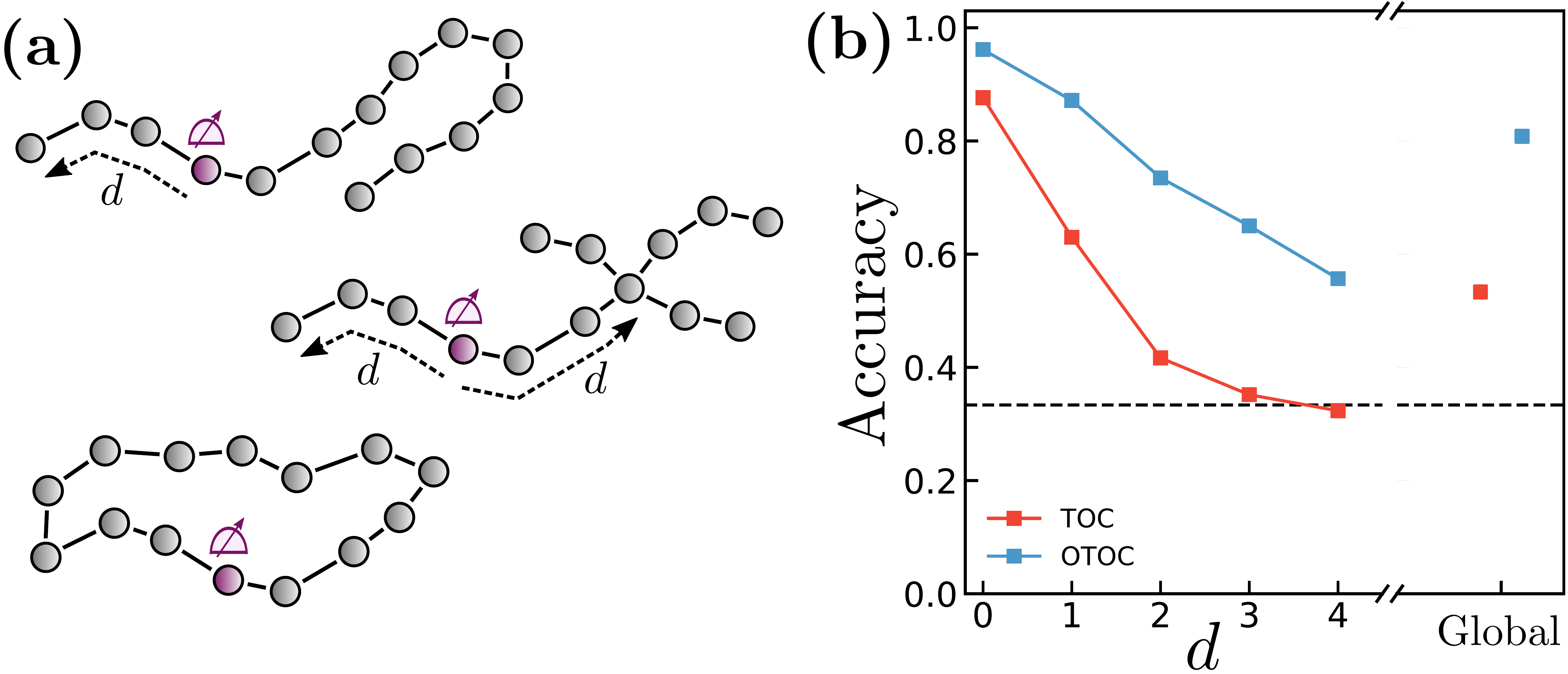}
\caption{
\textbf{(a)} The three spin geometries considered in the learning task defined in the text. Each geometry consists of $L=14$ spins. The probe qubit (purple) is located along a subset of the system that is identical between the three geometries up to a distance $d$ away from the probe.
\textbf{(b)} Accuracy of classification, using correlation functions that can be measured with (left) state preparation and read-out on the probe qubit and global unitary control over the remaining system, and (right) global state preparation, unitary control, and read-out.
For the former, the accuracy is plotted as a function of the distance $d$ of the probe qubit from the geometric feature of interest.
In both scenarios, access to OTOCs (blue) substantially improves the classification accuracy compare solely accessing TOCs (red).
} 
\label{fig: threeway}
\end{figure}

We now turn to learning when one has only global unitary control over the system of interest.
We consider a learning task where one wishes to classify the geometry of an unknown spin system, which we assume is drawn with equal probability from the three geometries shown in Fig.~\ref{fig: threeway}(a).
We find that access to OTOCs provides a substantial advantage in this classification task.
Notably, we find that OTOCs continue to improve learning even when one has only global state preparation, control, and read-out  (i.e.~even in the absence of a probe qubit).

The classification problem we consider is a close variant of those introduced in the main text.
We suppose that one has access to the correlation functions of an unknown Hamiltonian whose connectivity corresponds to one of the three geometries shown in Fig.~\ref{fig: threeway}(a).
The goal is to distinguish which geometry describes the Hamiltonian.
We again approach this task by training and testing a support vector machine on samples of disorder realizations, see Section~\ref{sec: learning details} for details.

We consider learning in two different experimental access scenarios.
First, we consider the scenario where one has state preparation and read-out from a single probe qubit, and global control over the remainder of the system.
In this case, we take the probe qubit to be a distance $d$ away from any distinguishing features of the geometry (see Fig.~\ref{fig: threeway}), and study the learnability as a function of $d$.
Note that we are restricted to relatively small distances, $d \leq 4$, owing to the particular form of the three geometries considered.
We find that access to OTOCs increases the classification accuracy between $10\%$ and $35\%$ for all values of $d$ [Fig.~\ref{fig: threeway}(b)].
For instance, OTOCs allow classification with accuracy $\sim 65\%$ at $d=3$, at which learning via TOCs has nearly trivial accuracy.

Our second scenario is even more restrictive: we suppose that one has only global state preparation, control and read-out over the entire system.
Despite being commonplace in experiments such as NMR spectroscopy~\cite{laws2002solid}, learning in this scenario remains quite difficult in strongly-interacting systems, due to the combination of time-ordered correlators decaying quickly and local information being averaged out by global control and measurement.
Indeed, in our learning task, we find that learning via TOCs features a classification accuracy of only $\sim \! 55\%$.
Intuitively, we expect access to global OTOCs to improve learning, as operator spreading at late times is dependent on global geometric features of the system.
In keeping with this intuition, we find that learning via both TOCs and OTOCs improves the classification accuracy to $\sim \! 80\%$.

\end{document}